\documentclass{ws-mpla}
\newcommand{\NP}[1]{{ Nucl.\ Phys.\ } {\bf  #1}}
\newcommand{\ZP}[1]{{ Z.\ Phys.\ } {\bf  #1}}

\newcommand{\PL}[1]{{ Phys.\ Lett.\ } {\bf  #1}}

\newcommand{\PR}[1]{{ Phys.\ Rev.\ } {\bf  #1}}
\newcommand{\PRL}[1]{{ Phys.\ Rev.\ Lett.\ } {\bf  #1}}

\newcommand{\lsim}{\raise.3ex\hbox{$<$\kern-.75em\lower1ex\hbox{$\sim$}}}
\newcommand{\ima}{{\mbox{Im}\,}}
\newcommand{\rea}{{\mbox{Re}\,}}
\newcommand{\be}{\begin{equation}}
\newcommand{\ee}{\end{equation}}
\begin{document}

\markboth{J.R. Pel\'aez}
{Light scalars as tetraquarks or two-meson states from large $N_c$ and
unitarized Chiral Perturbation Theory.}

%
\catchline{}{}{}{}{}
%

\title{Light scalars as tetraquarks or two-meson states from large $N_c$\\ and
unitarized Chiral Perturbation Theory.
}

\author{\footnotesize J.R. PELAEZ}

\address{Departamento de F\'{\i}sica Te\'orica II. Universidad Complutense.
28040 Madrid. SPAIN \\
jrpelaez@fis.ucm.es}

\maketitle

\pub{Received (Day Month Year)}{Revised (Day Month Year)}

\begin{abstract}
By means of unitarized Chiral Perturbation Theory it is possible
to obtain a remarkable description of meson-meson scattering amplitudes up to 1.2 GeV,
and generate poles associated to
scalar and vector resonances. Since Chiral Perturbation Theory is the QCD low
energy effective theory, it is possible then to study its large $N_c$
limit where $q\bar q$ states are easily identified. The
vectors thus generated follow closely a $q\bar q$ behavior, whereas the 
light scalar poles follow the large $N_c$ behavior expected for a dominant
tetraquark or two-meson structure.
\vskip .5 cm

\centerline{\it Invited ``Brief Review'' to appear in Modern Physics Letters A.}

\keywords{scalar mesons, chiral symmetry, large $N_c$}
\end{abstract}

\ccode{PACS Nos.: 12.39.Fe, 11.15.Pg, 12.39.Mk, 13.75.Lb}

\section{Introduction}

We review here how the spectroscopic nature of the
lightest scalar resonances can be obtained 
from the large $N_c$ behavior\cite{Pelaez:2003dy}
of their associated poles\cite{Coimbra} 
in the meson-meson scattering  
one-loop Chiral Perturbation Theory 
unitarized\cite{GomezNicola:2001as} amplitudes.

The existence and nature
of the lightest scalar resonances is a longstanding
 controversial
issue  for hadron spectroscopy.
Their relevance
 is twofold: First, one of the
most interesting features of QCD is its non-abelian nature, so that 
gluons interact among themselves and could
produce glueballs,
which are isoscalars. Moreover,
the lightest ones are
expected to be also scalars. Naively, once all 
quark multiplets are identified in the scalar-isoscalar sector,
whatever remains are good candidates for glueballs. Unfortunately,
the whole picture is more messy due to mixing, 
since the resonances we actually see are a superposition of 
different states. Second, it is also understood that
QCD has an spontaneous breaking of the chiral symmetry since its
vacuum is not invariant under chiral transformations. The study
of the scalar-isoscalar sector is relevant to understand the QCD vacuum,
which has precisely those quantum numbers.

Although QCD is firmly established as the fundamental theory
of strong interactions and its predictions have been 
thoroughly tested to 
great accuracy
in the perturbative regime (above 1-2 GeV), it becomes non-perturbative at low energies, 
and helps little to address the existence and nature
of light scalar mesons. 
Hence, the usual perturbative expansion 
in terms of quarks and gluons has 
to be abandoned
in favor of somewhat less systematic approaches in terms of 
mesons. As a matter of fact, most chiral descriptions of meson dynamics 
do not include quarks and gluons and are hard to relate to QCD,
and the spectroscopic nature is thus imposed from the start. 
In contrast, 
models with quarks and gluons, even those
inspired in QCD, have problems with chiral symmetry, small
meson masses, etc... 

An exception
is the formalism of Chiral Perturbation Theory\cite{chpt1,chpt2} (ChPT), which is
the most general effective Lagrangian made out of pions, kaons and etas,
that respects the QCD chiral symmetry breaking pattern.
 These particles 
are the QCD low energy degrees of freedom
since they are Goldstone bosons of the QCD spontaneous
chiral symmetry breaking.
For meson-meson scattering  ChPT is an expansion in even
powers of momenta, $O(p^2), O(p^4)$...,
 over a scale $\Lambda_\chi\sim4\pi f_0\simeq 1\,$GeV.
Since the $u$, $d$ and $s$ quark
masses are so small compared with $\Lambda_\chi$ they
are introduced as perturbations, giving rise to the 
$\pi, K$ and $\eta$ masses, counted as $O(p^2)$. 
At each order, ChPT is the sum of {\it all terms}
 compatible with the symmetries,
multiplied by ``chiral'' parameters, that absorb
loop divergences order by order, yielding finite results.
ChPT is thus {\it the} quantum effective field theory of QCD,
and it allows for
a systematic {\it and model independent} analysis 
of low energy mesonic processes.

Since ChPT is an expansion in momenta and masses,
it is limited to low energies. 
As the energy grows, the ChPT truncated series
violate unitarity.
Nevertheless, in recent
years ChPT has been extended to higher energies by means of unitarization 
\cite{Dobado:1996ps,Dobado:1992ha,Oller:1997ng,Guerrero:1998ei,Oller:1997ti,JA,GomezNicola:2001as}. 
The results 
are remarkable\cite{GomezNicola:2001as}, extending
the one-loop ChPT calculations of two body $\pi$, K or $\eta$
scattering up to 1.2 GeV, but keeping simultaneously
the correct low energy expansion and with chiral parameters
compatible with standard ChPT. 
Furthermore it {\it generates} 
the poles associated to the $\rho(770)$, $K^*(892)$ and the octet  $\phi$
vector resonances, as well as those of the controversial scalar states,
namely: the $\sigma$, $\kappa$, $a_0(980)$, and the $f_0(980)$.

Among these states, the most controversial are
the $\sigma$, now called $f_0(600)$ 
in the Particle Data Group (PDG) Review \cite{PDG},  and the $\kappa$.
They appear as broad resonant structures 
in the scalar channels in meson-meson
scattering  since they
do not display a Breit-Wigner shape. Still, many 
groups\cite{Dobado:1992ha,Oller:1997ng,JA,newsigma,kappa} do find
an associated pole in the
amplitude, but deep in the complex plane. 
For a compilation of $\sigma$ and $\kappa$ poles 
see the nice overview in \cite{vanBeveren:2002vw}. 
Let us remark that meson-meson scattering data\cite{pipidata} 
are hard to obtain, since 
they are extracted from reactions
like meson-$N\rightarrow$meson-meson-$N$, but with assumptions
like a factorization of the 
four meson amplitude, or that only one meson is exchanged and that
it is more or less on shell, etc... All these approximations
introduce large systematic errors.
Very recently, however, other experiments on
meson-meson interactions have become available,
like the very precise
determination of a combination of $\pi\pi$ phase shifts
from $K_{l4}$ decays \cite{Pislak:2001bf}, or the
 results from charm decays \cite{charm}. The latter have analyzed the
pole structure of their amplitudes and seem to find
both the $\sigma$ and $\kappa$ poles in reasonable agreement
with the works mentioned above but in completely different processes.

The controversy about their spectroscopic nature is even stronger.
The interest of
unitarized ChPT is that these states are generated
from chiral symmetry and unitarity,
without any prejudice toward their existence or structure. 
Amazingly the nine scalars are
generated together, suggesting they 
form an SU(3) multiplet, with a similar composition, although 
probably mixed with other states when possible.

Once again,  ChPT has the advantage
that it is the QCD low energy effective theory, 
and should behave as such in certain limits.
For this reason we have studied\cite{Pelaez:2003dy} the 
large $N_c$ expansion\cite{'tHooft:1973jz},
which  is the only
analytic approximation to QCD in the whole
energy region. Remarkably, it
provides a clear definition of $\bar qq$ states, that become
bound states when $N_c\rightarrow\infty$, whereas tetraquark 
or two meson states do not.
Starting from the unitarized ChPT  
meson-meson scattering, and scaling the parameters with $N_c$
according to the QCD rules\cite{chptlargen}, it has been possible to show that
vector mesons follow remarkably well their $q\bar q$
expected behavior, whereas all scalar nonet candidates 
behave as if their main component
had a tetraquark or two-meson structure.

In the next two sections we review in 
more detail the description of meson-meson
scattering with unitarized ChPT. In Section 4 we 
determine the parameters of the poles in those amplitudes, 
and in Section 5 we will determine how
those poles behave in the large $N_c$ limit. 
Finally we will present some brief conclusions.

\section{Meson-meson scattering and Chiral Perturbation Theory}

Chiral Perturbation Theory\cite{chpt1,chpt2}  (ChPT) is
built as the most general derivative expansion of a Lagrangian
containing {\it only} pions, kaons and the eta. These particles 
are the Goldstone bosons associated to the spontaneous
chiral symmetry breaking of massless QCD.
In practice, and for meson-meson scattering,
ChPT becomes an expansion in even
powers of energy or momenta, denoted as $O(p^2), O(p^4)$, etc...,
over the scale of the spontaneous breaking, i.e.,$4\pi f_0\simeq 1.2\,$GeV.
Of course, quarks are not massless, but their masses
are small compared with typical hadronic scales, and they
are introduced as perturbations 
 in the same power counting, giving rise to the 
$\pi, K$ and $\eta$ masses, counted as $M=O(p^2)$. 
The main advantage of ChPT 
is that it provides a Lagrangian that
allows for true  Quantum Field Theory calculations, 
and a chiral power counting to organize {\it systematically}
the size of the corrections at low energies.
For example, in the isospin limit, the leading order Lagrangian is universal
since it only depends on $f_0$,
the pion decay constant in the chiral limit,
and the leading order masses $M^0_\pi, M^0_K$ and $M^0_\eta$.
However, it is possible to calculate
meson loops, whose divergences are renormalized in a finite set of
chiral parameters
at each order in the expansion.
The dependence on the QCD dynamics only comes through 
higher order parameters. For instance, meson-meson scattering
to one loop depends on eight $L_i$ parameters\cite{chpt2}, 
whose values are shown in Table 1.
As usual after renormalization, they
depend on a
 regularization scale $\mu$:
\begin{equation}
 L_i(\mu_2)=L_i(\mu_1)+\frac{\Gamma_i}{16\pi^2}\log\frac{\mu_1}{\mu_2},
\label{Lis}
\end{equation}
where $\Gamma_i$ are constants\cite{chpt1}.
Of course, in physical observables the $\mu$ dependence is canceled
through the regularization of the loop integrals. 
This procedure can be repeated  obtaining
finite results at each order.
As long as we remain at low energies, only a few
orders are necessary and the theory is
predictive, since once the set of parameters
up to that order is fixed from some experiments,
it {\it should} describe, to that order,
 any other 
physical process involving mesons. 

\begin{table}[hbpt]
\tbl{$O(p^4)$ chiral parameters ($\times10^{3}$) and their $N_c$ scaling.
In the ChPT column, $L_1,L_2,L_3$ come from$^{23}$  and  the rest from$^4$.
The IAM
columns correspond to different fits$^3$
}
{\begin{tabular}{|c||c||c||c|c|c|}
\hline
$O(p^4)$&$N_c$&ChPT&IAM I&IAM II&IAM III\\
Parameter& Scaling &$\mu=770\,$MeV&$\mu=770\,$MeV&$\mu=770\,$MeV&$\mu=770\,$MeV\\
\hline
$L_1$
& $O(N_c)$
& $0.4\pm0.3$
& $0.56\pm0.10$ 
& $0.59\pm0.08$
& $0.60\pm0.09$
\\
$L_2$
& $O(N_c)$
& $1.35\pm0.3$ 
& $1.21\pm0.10$ 
& $1.18\pm0.10$
& $1.22\pm0.08$\\
$L_3 $  &
 $O(N_c)$&
 $-3.5\pm1.1$&
$-2.79\pm0.14$ 
&$-2.93\pm0.10$
& $-3.02\pm0.06$
\\
$L_4$
& $O(1)$
& $-0.3\pm0.5$& $-0.36\pm0.17$ 
& $0.2\pm0.004$
& 0 (fixed)\\
$L_5$
& $O(N_c)$
& $1.4\pm0.5$& $1.4\pm0.5$ 
& $1.8\pm0.08$
& $1.9\pm0.03$
\\
$L_6$
& $O(1)$
& $-0.2\pm0.3$& $0.07\pm0.08$ 
&$0\pm0.5$
&$-0.07\pm0.20$\\
$L_7 $  & $O(1)$ & 
$-0.4\pm0.2$&
$-0.44\pm0.15$ &
$-0.12\pm0.16$&
$-0.25\pm0.18$
\\
$L_8$
& $O(N_c)$
& $0.9\pm0.3$& $0.78\pm0.18$ 
&$0.78\pm0.7$
&$0.84\pm0.23$\\
\hline
$2L_1-L_2$
& $O(1)$
& $-0.55\pm0.7$& $0.09\pm0.10$ 
&$0.0\pm0.1$
&$-0.02\pm0.10$\\
\hline
\end{tabular}}
\end{table}

 Other salient features of ChPT
are its model independence and that it has been
possible to 
obtain\cite{chpt2,chptlargen} from QCD
 the $L_i$ large $N_c$ behavior, given in Table 1. Also, 
these parameters contain information about heavier\cite{Ecker:1988te} meson states
that have not been included as degrees of freedom in ChPT.

However,  ChPT is limited to low energies, since
the number of terms allowed by symmetry
  increases dramatically  at each order, because 
the ChPT series violate unitarity
as the energy increases, and finally, due to
resonances that appear rather soon
in meson physics. These states are
associated to poles in the second Riemann sheet of the amplitudes
that cannot be accommodated by the 
series of ChPT, which are polynomial (there are also logarithmic terms, 
irrelevant for this discussion).

For the above reasons, in recent
years ChPT it has been extended to higher energies by means of unitarization 
\cite{Dobado:1996ps,Dobado:1992ha,Oller:1997ng,Guerrero:1998ei,Oller:1997ti,JA,GomezNicola:2001as}, that we discuss next.

\section{Unitarized Chiral Perturbation Theory}

In order to compare with experiment
it is customary to use partial waves $t_{IJ}$ of
definite isospin $I$ and angular momentum $J$. 
For simplicity we will omit 
the $I,J$ subindices in what follows, so that the chiral expansion becomes
$t\simeq t_2+t_4+...$, with $t_2$ and $t_4$ of ${O}(p^2)$ and ${O}(p^4)$, respectively. 
The unitarity relation  for the partial waves $t_{ij}$,
where $i,j$ denote the different available states, is very simple: 
when two states, say "1" and ``2'', are accessible,
it becomes
\be
\ima T = T \, \Sigma \, T^* \quad \Rightarrow \quad \ima T^{-1}=- \Sigma
\quad  \Rightarrow \quad T=(\rea T^{-1}- i \,\Sigma)^{-1}
\label{unimatrix}
\ee
with \vspace*{-.5cm}
\be
T=\left(
\begin{array}{cc}
t_{11}&t_{12}\\
t_{12}&t_{22} \\
\end{array}
\right)
\quad ,\quad
\Sigma=\left(
\begin{array}{cc}
\sigma_1&0\\
0 & \sigma_2\\
\end{array}
\right)\,,
\ee
where $\sigma_i=2 q_i/\sqrt{s}$ and $q_i$ is the C.M. momentum of
the state $i$. The generalization to more than two accessible
states is straightforward in this notation.
Note that, since $\ima T^{-1}$ is fixed by unitarity
{\it we only need to know the real part of the Inverse Amplitude}.
Note that Eq.(\ref{unimatrix}) is non-linear and cannot be satisfied 
exactly with a perturbative expansion like that of ChPT, although
it holds perturbatively, i.e,
\vspace*{-.2cm}
\begin{eqnarray}
\ima T_2 = 0, \quad \quad \ima T_4 = T_2 \, \Sigma
\, T_2^*\,+ {O}(p^6) . \label{pertuni}
\end{eqnarray}
The use of 
the ChPT expansion 
$\rea T^{-1}\simeq  T_2^{-1}(1-(\rea T_4) T_2^{-1}+...)$ 
in eq.(\ref{unimatrix}),
guarantees that we reobtain the ChPT low energy expansion
and that we are taking into account all the information
included in the chiral Lagrangians (both about $N_c$ and 
about heavier resonances).
In practice, all the powers of $1/f_0$ in the amplitudes
are rewritten 
in terms of physical constants $f_\pi$ or $f_K$ or $f_\eta$.
At leading order this difference is irrelevant, but at 
one loop, we have three possible choices for each power of 
$f_0$ in the amplitudes, all equivalent up to ${O}(p^6)$. 
It is however possible to substitute the $f_0$'s by
their expressions in terms of $f_\pi$ or $f_K$ or $f_\eta$
in such a way that they cancel the ${ O}(p^6)$ and higher order
contributions in eq.(\ref{pertuni}), so that
\vspace*{-.2cm}
\begin{eqnarray}
\ima T_2 = 0, \quad \quad \ima T_4 = T_2 \, \Sigma
\, T_2^*. \label{exactpertuni}
\end{eqnarray}
We will call these conditions ``exact perturbative unitarity''.
From  eqs.(\ref{unimatrix}),(\ref{exactpertuni}), and the $\rea T^{-1}$
ChPT expansion, we find
\begin{equation}
 T\simeq T_2 (T_2-T_4)^{-1} T_2,
\label{IAM}
\end{equation}
which is the coupled channel IAM
that has been used 
to unitarize simultaneously all the one-loop ChPT meson-meson
scattering amplitudes\cite{GomezNicola:2001as}. Since 
all the pieces are analytic it is straightforward
to continue analytically to the complex $s$ plane 
and look for poles associated to resonances. 
Indeed, the analytic continuation 
to the complex $s$ plane has also been justified in terms 
of dispersion relations in the elastic case\cite{Dobado:1996ps}. Alternative methods
have been proposed and applied successfully to the one loop ChPT, 
for instance for $\pi\pi$ scattering\cite{Nieves:1999bx}.
However, in this brief review
we concentrate just on the IAM 
due to its remarkable success and simplicity, since it only involves
algebraic manipulations on the ChPT series.

The IAM was applied first for 
partial waves in the elastic region, where a single channel
is enough to describe the data. 
This approach was able to generate\cite{Dobado:1996ps}.
the $\rho$ and $\sigma$ poles
in $\pi\pi$ scattering and that of  $K^*$ in $\pi K\rightarrow\pi K$. 
Later on, it has been noticed that the $\kappa$ pole is also
obtained within the single channel formalism.
Concerning coupled channels, since
only the the $\pi\pi$, $\pi K$ and $\pi \eta$ amplitudes\cite{chpt1,Kpi}
were known at that time, at first\cite{Oller:1997ng} only
the leading order and the dominant s-channel loops
were  considered,
neglecting crossed and tadpole loop
diagrams. Despite these approximations,
a remarkable description of meson-meson
scattering was achieved up to 1.2 GeV, generating  the poles associated to the 
$\rho$, $K^*$, $f_0$, $a_0$, $\sigma$ and $\kappa$.
The price to pay was, first,  that only 
the leading order of the expansion
was recovered at low energies. Second, apart from the fact
that loop divergences  were regularized with a cutoff, 
thus introducing an spurious parameter,
they 
were not completely renormalized, due to
the missing diagrams. Therefore, 
it was not possible to compare the $L_i$,
which encode the underlying QCD dynamics, 
with those already present in the literature.
In addition, the study of the large $N_c$ limit
is cumbersome due to the incomplete renormalization and
since we do not know how the cutoff should scale.

 As already explained
the whole approach is relevant to study the existence 
and properties of 
light scalars and 
it is then very important to check that
these poles and their features
are not just artifacts of the approximations,
estimate the uncertainties in their parameters, and
check their compatibility with other data. These considerations
triggered the interest
in calculating and unitarizing the remaining meson-meson amplitudes
within one-loop ChPT. Hence, 
the
$K\bar{K}\rightarrow K\bar{K}$ calculation
was completed\cite{Guerrero:1998ei}, thus
allowing for the unitarization of the $\pi\pi$, $K\bar{K}$
coupled system. There was a good agreement
of the IAM description with the existing $L_i$, reproducing
again the resonances in that system.
More recently, we have completed\cite{GomezNicola:2001as}  
the one-loop meson-meson scattering calculation,
including three new amplitudes:
$K\eta\rightarrow K\eta$, $\eta\eta\rightarrow\eta\eta$ and
$K\pi\rightarrow K\eta$, but recalculating
the other five amplitudes, unifying the  notation, ensuring
`` exact perturbative
unitarity'', Eq.(\ref{exactpertuni}), 
and also correcting some errors in the literature.
Next, we have unitarized these amplitudes with the coupled channel IAM, 
which allows for a direct comparison with the
standard low-energy chiral parameters, in very good
agreement with previous determinations. 
In that work 
we presented the full calculation of all the one-loop
amplitudes in dimensional regularization, 
and a 
simultaneous description of
  the low energy and the resonance regions.

The first check was to use the  
standard ChPT parameters, given in Table 1
to see if the resonant features were still there, at least qualitatively,
and they are. Thus, they are not just an artifact of the approximations 
and of the values chosen for the parameters.
As already commented, this comparison can only be performed now since
we have all the amplitudes
renormalized in the standard $\overline{MS}-1$ scheme.

After checking that, we made an IAM fit.  
Systematic uncertainties are the largest contribution to the
resulting error bands 
as well as in the fit parameters in Table 1.
These error bands
are calculated from a MonteCarlo Gaussian sampling\cite{GomezNicola:2001as}
(1000 points)
of the $L_i$ sets within their uncertainties, assuming they are
uncorrelated. Note that in Table 1 we list three sets of
parameters for the IAM fit, fairly compatible among them and with 
those of standard ChPT. 
They correspond to different choices when reexpressing
the $f_0$ parameter of the Lagrangian in terms of
physical decay constants.
The IAM I fit was obtained\cite{GomezNicola:2001as} using just $f_\pi$,
which is simpler but unnatural 
when dealing with kaons or etas.  
There, it could be observed that the $f_0(980)$  region
was not very well described yielding a
too small width for the resonance.

\begin{figure}[hbpt]
\centerline{\psfig{file=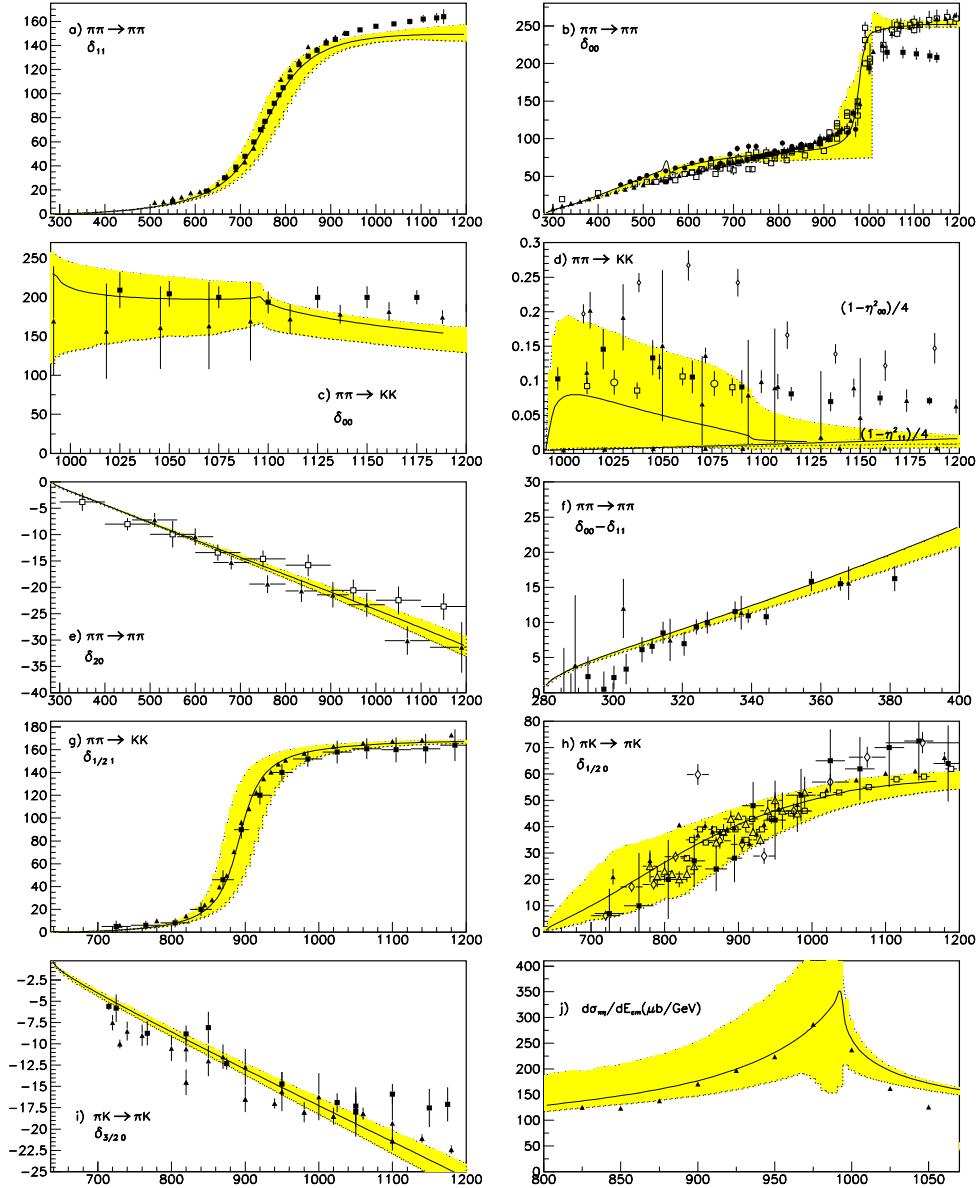,width=5.0in}}
\caption
{IAM fit to meson-meson scattering data, set II in Table 1.
The uncertainty bands are dominated by systematic errors in the data$^{17}$.
}
\label{fig:allIAM}
\end{figure}

For that reason, Fig.1 shows the results of
a second fit\cite{Coimbra} (IAM II) using
amplitudes written in terms of 
$f_K$ and $f_\eta$ when dealing with processes
involving kaons or etas.
Let us remark that the data in the $f_0(980)$ region
is well within the uncertainties.
In particular, in 
the $O(p^2)$ one factor of $1/f_\pi$ has been replaced\cite{Coimbra} by
$1/f_K$ for each two kaons present
between the initial or final state, or by $1/f_\eta$ 
for each two etas appearing between the initial and final states.
In the special case $K\eta\rightarrow K\pi$ 
$1/f_\pi^2$ has been changed by $1/(f_Kf_\eta)$.
The difference between the two ways of writing
the leading order amplitudes is $O(p^4)$, and is therefore included
in the next to leading order contribution using the relations\cite{chpt2,GomezNicola:2001as}
 between the decay constants and $f_0$. 
The $1/f_0$ factors in each loop function  at $O(p^4)$
(generically, the $J(s)$
given in the appendix of \cite{GomezNicola:2001as})
are changed to satisfy ``exact perturbative unitarity'',
eqs.(\ref{exactpertuni}). According to 
the ChPT counting, the amplitudes are the same 
up to $O(p^4)$, but numerically they are slightly different.
A similar choice has been 
suggested\cite{Descotes-Genon:2003cg} independently within
non-unitarized standard SU(3) ChPT to avoid
the uncertainties arising from fluctuation of vacuum $s\bar  s$
pairs. In particular, it is suggested to obtain
elastic amplitudes $A_{PQ}$ from the safe combinations
$F_\pi^4 A_{\pi\pi}$ or $F_\pi^2F_K^2 A_{\pi K}$, etc...
For external fields this amounts to our choice, and the
normalization of internal loop function is then dictated by exact perturbative
coupled channel unitarity. 
From Table 1, we see that 
the only sizable change is in the parameters related to the 
decay constants, i.e., $L_4$ and $L_5$. 
For illustration we give in Table 1 a third fit, IAM III, 
obtained as IAM II but fixing $L_4=0$, 
its large $N_c$ limit, as it is done in
recent $K_{l4}$  $O(p^4)$ 
determinations. 
 As seen in Table 1, these 
chiral parameters are compatible with those from standard ChPT,
which means that we have a simultaneous description of the low energy and resonance
regime.
Finally, since the expressions
are fully renormalized, all the QCD $N_c$ dependence appears correctly
through the $L_i$ 
and cannot  hide in any spurious parameter.

At this point one might wonder about
higher order effects. There is a simple way to 
extend the IAM to higher orders\cite{Dobado:1996ps},
first applied to two loop $\pi\pi$ scattering and the
pion form factor\cite{Hannah:1997ux},
with remarkable results.
This study has been extended\cite{Nieves:2001de} 
with a careful analysis of the uncertainties. The amplitude depends
on the  $O(p^4)$ and $O(p^6)$ parameters through six combinations, 
called $b_i$.
Despite the poor knowledge about these two-loop parameters
a good description of the data is found\cite{Nieves:2001de}, 
including the $\sigma$ and $\rho$ regions,
with parameters compatible within errors with those 
in the literature. The error analysis\cite{Nieves:2001de} 
 is also of relevance because it was shown that 
the IAM
crossing violations, {\it taking into account
the present
experimental uncertainties}, are `` not very large in percentage terms''.

\section{Poles associated to resonances}

The most interesting feature of chiral unitary
approaches is that the poles thus generated
are not included in the original ChPT Lagrangian and hence
appear without theoretical prejudices toward
their existence, classification in multiplets, or nature.
Remarkably,
the scalar resonances $\sigma, \kappa, a_0(980)$ and $f_0(980)$
appear together
in those chiral unitarized amplitudes, and it seems
natural to interpret them as a nonet (see also\cite{Black:1998wt}).
Nevertheless, we should distinguish two 
different resonance generation mechanisms:
it had already been noted\cite{JA}, that to generate
the scalars just the leading order and
a cutoff was enough, whereas the vector mesons
require the chiral parameters, 
particularly\cite{Dobado:1992ha,Oller:1997ng} $L_1, L_2$ and $L_3$. 
Of course, the chiral parameters are always present,
but changes within errors in their values do not affect
the existence of the light scalar poles, but just the details of their description.
Since the vectors are  fairly well established $q\bar{q}$ states,
this difference suggests that scalars like the $\sigma$, $\kappa$, etc,
may have a different nature. With the amplitudes described in the previous
sections we expect to reach a more conclusive statement.

Thus, in Table 2 we show the pole position for the resonances\cite{Coimbra}, 
including  uncertainties,
for the different IAM parameter sets given in Table 1.
For comparison we also provide those obtained
in the ``approximated'' IAM\cite{Oller:1997ng}, 
whereas we list in Table 3 the poles listed presently in the
PDG\cite{PDG}. These results deserve some comments:
\begin{itemlist}

\item The vectors $\rho(770)$ and $K^*(892)$, are very stable
within chiral unitary approaches. Their positions are almost
the same irrespective of whether one uses the single 
channel\cite{Dobado:1996ps}, 
the approximated coupled channel\cite{Oller:1997ng}
or the complete IAM\cite{GomezNicola:2001as}.

\item The $\sigma$ and $\kappa$ pole positions are robust
within these approaches. No matter what version
of the IAM is used. Note the small uncertainties in some of their parameters,
in very good agreement with recent experiments\cite{charm}.

\item  The $f_0(980)$  has  a sizable
 decay to two different channels and therefore it can only be studied
in a coupled channel formalism. In practice, both the 
approximated and complete IAM generate a pole
associated to this state at approximately the same mass.
However, as already remarked, the unitarized ChPT amplitudes 
using just\cite{GomezNicola:2001as} $f_\pi$, yield a 
too narrow width. The good news is that it can be well 
accommodated\cite{Coimbra} using $f_\pi$, $f_K$ and $f_\eta$.

\item   The $a_0(980)$ also requires a coupled channel formalism,
and the data on this region is well described either by the approximated or
the complete IAM. However, the presence of a pole is 
strongly dependent on whether we write the ChPT amplitudes
only in terms of $f_\pi$, or using the three $f_\pi$, $f_K$ and $f_\eta$.
In the first case, it has been pointed out\cite{Uehara:2002nv},
that the use of the
``approximated'' IAM with just $f_\pi$, favored 
a ``cusp'' interpretation
of the $a_0(980)$ enhancement in $\pi\eta$ production. 
With the complete IAM we do not find a pole near the
$a_0(980)$ enhancement and indeed the $\pi\eta$ phase-shift
does not cross $\pi/2$ and it neither has a fast phase movement.
In contrast, when expressing $f_0$ in terms of
 $f_\pi$, $f_K$ and $f_\eta$ as described in the previous section,
we do find a pole and its associated fast phase movement through $\pi/2$,
either with the approximated or complete IAM. 
Thus, this pole is rather unstable as can 
be noticed from its large uncertainties in Table 2. 

\item We are more
favorable toward the pole interpretation for the $a_0(980)$
because, first, it is also able to 
describe better the $f_0(980)$ width. Second, we have already remarked
that even within standard SU(3) ChPT, the scattering amplitude calculations
are safer against $s\bar s$ vacuum fluctuations  if these fluctuation terms,
related to $L_4$ and $L_6$
are absorbed\cite{Descotes-Genon:2003cg} in terms of physical 
decay constants $f_\pi, f_K$ and $f_\eta$ when expanding the anplitudes.
\end{itemlist}

\begin{table}[htbp]
\tbl{ Pole positions (with errors) in meson-meson scattering.
When close to the real axis the mass and width of the 
associated resonance is $\sqrt{s_{pole}}\simeq M-i \Gamma/2$.}
{\setlength{\tabcolsep}{1.7mm}
\begin{tabular}{ccccccc}
\hline
$\sqrt{s_{pole}}$(MeV)
&$\rho$
&$K^*$
&$\sigma$
&$f_0$
&$a_0$
&$\kappa$
\\ \hline
$^{\hbox{IAM Approx}}_{\;\;\;
\hbox{(no errors)}}$
&759-i\,71
&892-i\,21
&442-i\,227
&994-i\,14
&1055-i\,21
&770-i\,250
\\\hline
IAM I
&760-i\,82
&886-i\,21
&443-i\,217
&988-i\,4
& cusp?
&750-i\,226
\\
(errors)
&$\pm$ 52$\pm$ i\,25
&$\pm$ 50$\pm$ i\,8
&$\pm$ 17$\pm$ i\,12
&$\pm$ 19$\pm$ i\,3
&
&$\pm$18$\pm$i\,11
\\ \hline
IAM II
&754-i\,74
&889-i\,24
&440-i\,212
&973-i\,11
&1117-i\,12
&753-i\,235
\\
(errors)
&$\pm$ 18$\pm$ i\,10
&$\pm$ 13$\pm$ i\,4
&$\pm$ 8$\pm$ i\,15
&$^{+39}_{-127}$ $^{+i\,189}_{-i\,11}$
&$^{+24}_{-320}$ $^{+i\,43}_{-i\,12}$
&$\pm$ 52$\pm$ i\,33\\\hline
IAM III
&748-i68
&889-i23
&440-i216
&972-i8
&1091-i52
&754-i230
\\
(errors)
&$\pm$ 31$\pm$ i\,29
&$\pm$ 22$\pm$ i\,8
&$\pm$ 7$\pm$ i\,18
&$^{+21}_{-56}$$\pm$ i\,7
&$^{+19}_{-45}$ $^{+i\,21}_{-i\,40}$
&$\pm$ 22$\pm$ i\,27\\
\hline
\end{tabular}
}  
\vskip .5cm

\tbl{ Mass and widths or pole positions 
of the light resonances quoted in the PDG$^{12}$
Recall that for narrow resonances $\sqrt{s_{pole}}\simeq M-i \Gamma/2$}
{
\setlength{\tabcolsep}{1.1mm}
\begin{tabular}{ccccccc}
\hline
PDG2004
&$\rho(770)$
&$K^*(892)^\pm$
&$\sigma$ or $f_0(600)$
&$f_0(980)$
&$a_0(980)$
&$\kappa$
\\ \hline
Mass (MeV)
&$775.8\pm0.5$
&$891.66\pm0.26$
&(400-1200)-i\,(300-500)
&$980\pm10$
&$984.7\pm1.2$
&not\\
Width (MeV)
& $150.3\pm1.6$
& $50.8\pm0.9$
&(we list the pole)
&40-100
&50-100
&listed\\\hline
\end{tabular}
}  
\end{table}

Let us remark that the $f_0(980)$ and $a_0(980)$ resonances
are very close to the $K\bar{K}$ threshold, which can induce a considerable
distortion in the resonance shape, whose relation to the pole position
could be far from that expected for narrow resonances. In addition
these states have a large mass and it is likely that their nature 
should be understood from a mixture with heavier states.

\section{Resonance Behavior in the large $N_c$ limit}

First of all, let us recall
that QCD not only predicts 
$\bar q q$ states to become bound states 
in the $N_c\rightarrow\infty$ limit, but
it tells us exactly {\it what is the large} $N_c$ 
{\it dependence of their mass and width}: their mass should remain constant,
whereas their width vanishes as $1/N_c$.
For instance, the $\pi,K,\eta$ masses behave as $O(1)$ and 
$f_0$, their decay constant in the chiral limit, as $O(\sqrt{N_c})$.

The $N_c$ scaling
of the $L_i$ parameters\cite{chpt1,chptlargen} is listed in Table 1.
Let us then scale $f_0\rightarrow f_0 \sqrt{N_c/3}$
and $L_i(\mu)\rightarrow L_i(\mu)(N_c/3)$ for $i=2,3,5,8$, keeping
the masses, $2L_1-L_2,L_4,L_6$ and $L_7$ constant. We use IAM set II
in Table 1 for the parameters at $N_c=3$.
In Fig. 2, we show how the poles, represented by a dot,
 move in the lower half of the complex plane as $N_c$ changes.
On top, we see the $\rho(770)$ and $K^*(892)$ vector mesons,
whose poles move toward the real axis.
That is, the vectors widths vanish at large $N_c$, thus becoming 
bound states.
In contrast, in the bottom, we see that
both the $\sigma$ and $\kappa$ poles move {\it away} from the real axis, which
means that they dissolve in the continuum.
\begin{figure}[hbpt]
\centerline{\psfig{file=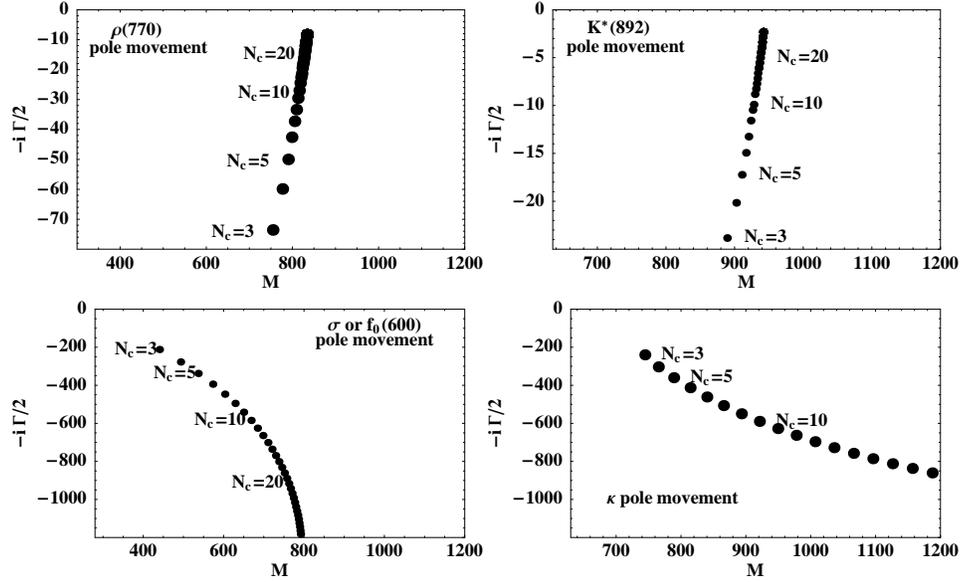,width=5.0in}}
\caption{Large $N_c$ behavior of poles
in the lower half of the second Riemann 
sheet of the unitarized one-loop meson-meson scattering amplitude
from ChPT.
For each value of $N_c$ the pole is represented by a dot,
in different meson-meson
scattering channel. 
Note that the $\sigma$ and $\kappa$ behavior is opposite to that
of well know vector states as the $\rho$ and $K^*$.
}
\label{fig:fln}
\end{figure}

Not only we can see that vectors become bound states and scalars do not,
but we can determine what is the precise $N_c$ dependence.
Before doing that, we should nevertheless recall that
it is not known at what scale $\mu$ to apply the large $N_c$ 
to the $L_i$. Certainly, the logarithmic
part in eq.(\ref{Lis}) is subleading, but
 it has been pointed out\cite{Pich:2002xy} that it
can be rather large for $N_c=3$,
which is our starting point for the $N_c$ evolution.
In addition, the scale dependence is 
suppressed by $1/N_c$ for $L_i=L_2,L_3,L_5,L_8$, but not for
$2L_1-L_2, L_4,L_6$ and $L_7$. 
The choice $\mu$, or, in other words, the choice of
``initial values'', is a systematic uncertainty
typical of the large $N_c$ approach.
Customarily, the uncertainty in the $\mu$ 
where the $N_c$ scaling applies is estimated\cite{chpt1} 
varying $\mu$ between 0.5 and 1 GeV.
We will check that this estimate is correct with the vector mesons,
firmly established as $\bar qq$ states.
In addition, and in order to show that the choice of IAM sets in Table 1
is irrelevant for our analysis, we change now to set III. The behavior
is exactly that already found in Fig.2 and in previous 
works\cite{Pelaez:2003dy,Coimbra}.

Let us then look back at vector mesons.
In Fig.3 we show, for increasing $N_c$, the modulus of the 
$(I,J)=(1,1)$ and $(1/2,1)$ amplitudes
with the Breit-Wigner shape of the $\rho$
and $K^*(892)$, respectively.
There is always a peak at an almost constant position,
becoming 
narrower as $N_c$ increases.
We also show the evolution
of the $\rho$ and $K^*$ pole positions,
related to their mass and width as $\sqrt{s_{pole}}\simeq M-i \Gamma/2$.
We have normalized both $M$ and $\Gamma$
to their value at $N_c=3$ in order to compare
with the expected $\bar{q} q$ behavior:
$M_{N_c}/M_3$ constant and  $\Gamma_{N_c}/\Gamma_3\sim 1/N_c$
The agreement is remarkable within
the gray band that covers the uncertainty 
$\mu=0.5-1\,$GeV where to apply the large $N_c$ scaling.
We have checked that outside that band, the behavior starts 
deviating from that of $\bar qq$ states, which confirms that
the expected scale range where the large $N_c$ scaling 
applies is correct.
\begin{figure}[hbpt]
\centerline{\psfig{file=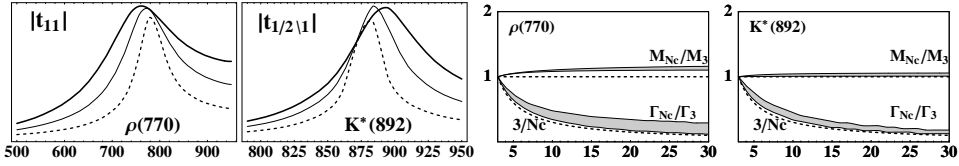,width=5.0in}}
\caption
{Left: Modulus 
of $\pi\pi$ and $\pi K$ elastic amplitudes versus $\sqrt{s}$
for $(I,J)=(1,1),(1/2,1)$:  
$N_c=3$ (thick line), $N_c=5$ (thin line) and $N_c=10$ 
(dotted line), scaled at $\mu=770\,$MeV.
Right: $\rho(770)$ and $K^*(892)$ pole positions: $\sqrt{s_{pole}}\equiv M-i\Gamma/2$
versus $N_c$. The gray areas cover the uncertainty $N_c=0.5-1\,$GeV. The dotted lines
show the expected $\bar q q$ large $N_c$ scaling.}
\label{fig:f1}
\end{figure}
\begin{figure}[hbpt]
\centerline{\psfig{file=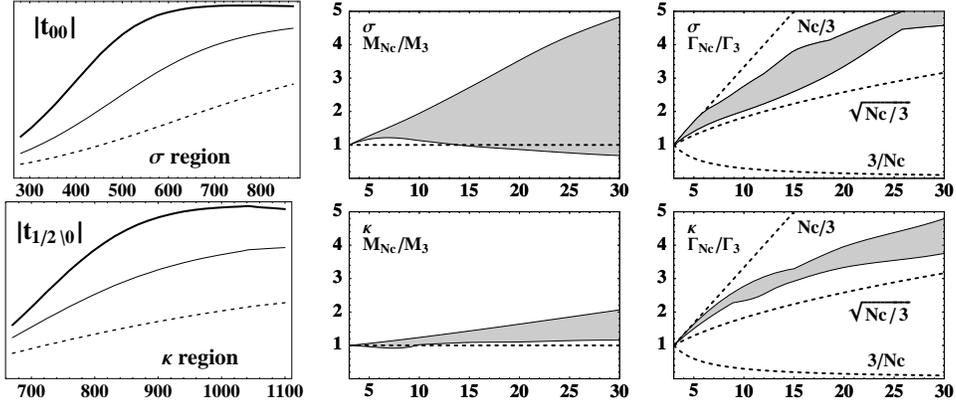,width=5.0in}}
\caption
{Top) Left:
 Modulus of the $(I,J)=(0,0)$ amplitude, versus $\sqrt{s}$
for $N_c=3$ (thick line), $N_c=5$ (thin line) and $N_c=10$ 
(dotted line), scaled at $\mu=770\,$MeV. Center: $\sigma$ mass
$N_c$ behavior. Right: $\sigma$ width ¡$N_c$ behavior. 
Bottom: The same but for the $(1/2,0)$ amplitude
and the $\kappa$.
}
\label{fig:f2}
\end{figure}

In Fig.4, in contrast, all over the $\sigma$ and $\kappa$ regions
the $(0,0)$ and $(1/2,0)$ amplitudes decrease as $N_c\rightarrow\infty$.
Their associated poles 
show a totally
different behavior, since 
{\em their width grows with} $N_c$, in
conflict with a ${\bar qq}$ interpretation. 
(We keep the $M$, $\Gamma$ notation, but now as definitions).
This is also suggested using the ChPT leading order
unitarized amplitudes with a 
regularization scale\cite{JA,Harada:2003em}.

In order to determine their spectroscopic nature,
we abandon momentarily the poles and look at the amplitudes
in the real axis and the representative Feynman diagrams given in 
in Fig. 5.
Imaginary parts are generated from s-channel intermediate 
states, as in Fig.5(a) or 5(c), when they are physically accessible. 
If there was a  
$\bar qq$ meson, with mass $M\sim O(1)$ and $\Gamma\sim 1/N_c$, 
we would expect at precisely $\sqrt{s}\simeq M$ 
that $\ima t\sim O(1)$ and a peak in the modulus, as it is indeed
the case of the $\rho$ and $K^*$. 
However, we have checked that {\em in the whole} $\sigma$ and $\kappa$ regions,
$\ima t\sim O(1/N_c^2)$ and $\rea t\sim O(1/N_c)$,
which means that
from $\bar q q$ states, the $\sigma$ and $\kappa$  
can only get real contributions from $\rho$ or $K^*$ t-channel exchange, 
respectively, as in Fig.5(b). 
The leading s-channel contribution for the $\kappa$ 
and $\sigma$ comes from Fig.5(c).
\begin{figure}[hbpt]
\centerline{\psfig{file=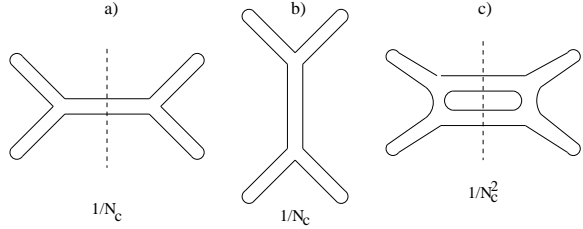,width=3.0in}}
\caption
{ Representative diagrams contributing to meson-meson scattering
and their $N_c$ scaling.}
\label{fig:diagrams}
\end{figure}

Then, one can immediately interpret
the $\kappa$ as a $\bar q \bar q qq$ (or two
meson) state, which  is predicted\cite{Jaffe} to unbound and  
become the meson-meson continuum when $N_c\rightarrow\infty$.
However, in the large $N_c$ limit, $\bar q \bar q qq$ 
and glueball exchange count the same, and our $N_c$ argument
{\it alone} cannot decide between both structures for the $\sigma$.
Nevertheless, given the fact that
glueballs are expected to have masses above 1 GeV, and that
the $\kappa$, which has strangeness and thus
cannot mix with a glueball,
 is a natural $SU(3)$ partner of the $\sigma$,
a dominant $\bar q \bar q qq$  or two-meson component for the $\sigma$ seems the
most natural interpretation, although most likely with some glueball
mixing.

Let us finally turn to the other members of the hypothesized scalar nonet:
the $f_0(980)$ and the $a_0(980)$. These resonances 
are more complicated due to the distortions 
caused by the nearby $\bar KK$ threshold, and poles
are harder to follow. 
Still, by looking in Fig.6 at the modulus of the amplitude $(0,0)$ 
in the vicinity of the $f_0(980)$,
we see that the characteristic sharp dip of the $f_0(980)$ 
vanishes when $N_c\rightarrow\infty$, at variance with
a $\bar qq$ state. For $N_c>5$ it follows again the $1/N_c^2$ 
scaling compatible 
with $\bar q\bar q qq$ states or glueballs.
The $a_0(980)$ behavior, shown in Fig.7, is more complicated.
When we apply the large $N_c$ scaling
at $\mu=0.55-1\,$ GeV, its
 peak disappears, suggesting that this is not a $\bar q q$
state, and $\ima t_{10}$
follows roughly the $1/N_c^2$ behavior
in the whole region. 
However, as shown in Fig.5, the peak does not vanish at
large $N_c$ if we take $\mu=0.5\,$GeV. Thus we cannot rule
out a possible $\bar q q$ nature, or a sizable mixing with $\bar q q$,
although it shows up in
an extreme corner of our uncertainty band.
 
Before concluding we want to remark that similar results have been 
obtained\cite{Uehara:2003ax}  for all resonances and just for central values of
the $L_i$ plus a cutoff, using the
approximated IAM\cite{Oller:1997ng}.
Furthermore, either the tetraquark structure of
these light scalar states, or the
fact that the lightest $\bar q q$ states are above $\sim$1 GeV,
 has received further support from
other large $N_c$ studies\cite{Harada:2003em,Cirigliano:2003yq} as
well as other kind of analysis\cite{other}.
\begin{figure}[hbpt]
\centerline{\psfig{file=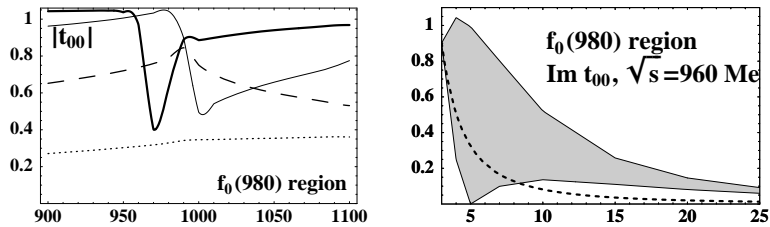,width=4.0in}}
\caption
{Left:
 Modulus of a $(I,J)=(0,0)$ amplitude, versus $\sqrt{s}$,
for $N_c=3$ (thick), $N_c=5$ (thin), $N_c=10$ 
(dashed) and $N_c=25$ 
(dotted), scaled at $\mu=770\,$MeV. Right: $\ima t_{00}$ 
versus $N_c$.}
\label{fig:f3}
\end{figure}
\begin{figure}[hbpt]
\centerline{\psfig{file=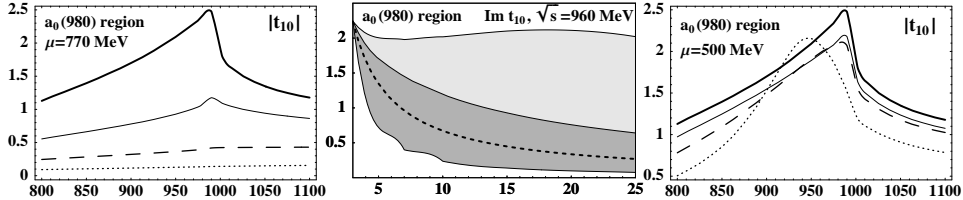,width=5.0in}}
\caption
{Left:
 Modulus of a $(I,J)=(1,0)$ scattering amplitude, versus $\sqrt{s}$,
for $N_c=3$ (thick), $N_c=5$ (thin), $N_c=10$ 
(dashed) and $N_c=25$ 
(dotted), scaled at $\mu=770\,$MeV. Right: scaled at $\mu=500\,$MeV.
Center: $\ima t_{00}$ versus $N_c$. The dark gray area covers 
the uncertainty $\mu=0.55-1\,$GeV, the light gray area from $\mu=0.5$ to $0.55\,$GeV.}
\label{fig:f4}
\end{figure}
\section{Conclusions}

We have reviewed a recent  set of works\cite{Pelaez:2003dy,Coimbra,GomezNicola:2001as}
in which we show how the unitarized one-loop Chiral Perturbation Theory (ChPT)
amplitudes generate poles associated to the 
lightest vector and scalar resonances and we study
their large $N_c$ behavior. This amplitudes respect the
chiral expansion and are fully renormalized. Indeed
they provide a remarkable description of two body scattering
of pions, kaons and etas up to 1.2 GeV using just the 
$O(p^4)$ ChPT parameters, with values compatible with previously 
existing determinations in the literature.

We have then studied\cite{Pelaez:2003dy} the evolution of the 
poles, mass and width associated to each one of these resonances,
through the QCD large $N_c$ scaling inherited
by the ChPT parameters.
We have found that the $\rho$ and $K^*$ vector mesons
follow remarkably well their expected $\bar q q$ behavior, both
qualitatively and quantitatively.
In contrast, the $\sigma$, $\kappa$, $f_0(980)$ and $a_0(980)$ 
large $N_c$ behavior
is in conflict with a $\bar qq$ nature
(not so conclusively for the $a_0(980)$), 
and strongly suggests a 
$\bar q \bar qqq$ or two meson main component, maybe with some mixing
with glueballs, when possible.

\section*{Acknowledgments}
I thank A. Andrianov, D. Espriu, A. G\'omez Nicola, F. Kleefeld,
R. Jaffe, E. Oset, J. Soto and  M. Uehara for their comments
and support from
the Spanish CICYT projects,
BFM2000-1326, BFM2002-01003 and the 
E.U. EURIDICE network contract no. HPRN-CT-2002-00311.

\end{document}